\documentclass[aps,prc,preprint,amsfonts,amsmath,superscriptaddress,floatfix,nofootinbib]{revtex4-2}
\usepackage{graphicx,amsmath,epsfig}
\newcommand{\beqy}{\begin{eqnarray}}
\newcommand{\eeqy}{\end{eqnarray}}

\newcommand{\rb}{\pmb{r}}
\newcommand{\rp}{\pmb{r^\prime}}

\newcommand{\sg}{\sigma}

\begin{document}

\title{Superfluid fraction in the crystalline crust of a neutron star: role of quantum zero-point motion of ions}

\author{N. Chamel}

\affiliation{Institut d'Astronomie et d'Astrophysique, CP-226, Universit\'e Libre de Bruxelles, 
1050 Brussels, Belgium}

\begin{abstract}
The suppression of the neutron superfluid fraction in the inner crust of a cold neutron star is 
mitigated by the quantum zero-point motion of ions about their equilibrium position. 
In turn, the crustal dynamics is altered by the presence of the neutron superfluid. These 
effects are studied self-consistently to assess the validity of the usual assumption of a 
perfect rigid lattice. To this end, fully three-dimensional band-structure calculations 
of the superfluid fraction are carried out in the weak-coupling approximation, considering 
body- and face-centered cubic lattices. In both cases, the superfluid fraction is still 
found to be strongly suppressed in the intermediate region of the inner crust. In turn, 
the effective mass of the ions is dramatically increased, thus further damping the 
ion fluctuations. These results are of relevance for the rotational and thermal evolutions 
of neutron stars. 
\end{abstract}


\maketitle

\section{Introduction}

Both microscopic calculations and astrophysical observations provide very strong evidence for 
the existence of a neutron superfluid in the inner crust of neutron stars (see, e.g., Refs.~\cite{chamel2017,haskell2018}
and references therein).  Due to the presence of spatial inhomogeneities,  the neutron superfluid density 
$\rho_{n,s}$ relating the spatially averaged neutron superfluid velocity $\pmb{\bar V_{n,s}}$ to the 
average neutron mass current $\pmb{\bar \rho_{n}}=\rho_{n,s}\pmb{\bar V_{n,s}}$ in the  crust frame 
is expected to 
be reduced compared to the mass density $\rho_{n,f}$ of free neutrons, as first studied in 
Refs.~\cite{chamel2004,CCH05a,chamel2005,CCH05b} (see also Ref.~\cite{chamel2017b} for a review).  The suppression of the superfluid fraction $\rho_{n,s}/\rho_{n,f}$ 
can be understood from the Bragg scattering of free neutrons by the crystal lattice. 

A similar phenomenon has been recently measured in the laboratory using cold atoms~\cite{chauveau2023,tao2023} 
thus confirming much earlier theoretical predictions~\cite{leggett1970} (see also Ref.~\cite{leggett1998}).  There is 
however an important difference: in these experiments the lattice is induced by external laser fields whereas 
in neutron stars the crustal lattice consists of positively-charged neutron-proton clusters formed from the spontaneous breaking 
of the translational symmetry of dense matter (these ions are immersed in a charge-compensating background of free 
electrons).  Previous calculations of $\rho_{n,s}$ assumed a perfect 
rigid lattice (see, e.g., Refs.~\cite{kashiwaba2019,sekizawa2022,minami2022,kenta2024,almirante2024,almirante2024b,chamel2025} for recent calculations).  However,  
it is well-known from X-ray diffraction by ordinary crystals that the quantum and thermal fluctuations of ions 
about their equilibrium position leads to the attenuation of the intensity of the diffracted beam (see, e.g., 
Ref.~\cite{kittel}). Kobyakov and Pethick~\cite{kobyakov2013} speculated that these fluctuations could have 
important implications for the dynamics of the neutron superfluid in the inner crust of neutron stars. But no
detailed calculations were performed. From the 
previous considerations, the superfluid fraction $\rho_{n,s}/\rho_{n,f}$ is expected to be increased by the lattice 
dynamics. In turn, this can affect the thermal and rotational evolutions of neutron stars (see, e.g., Refs.~\cite{andersson2012,chamel2013,chamelpagereddy2013,delsate2016}).

This paper is the second in a series. After investigating the role of BCS pairing on the neutron superfluid 
density in the inner crust of a cold neutron star~\cite{chamel2025}, the assumption of a rigid lattice and the 
influence of the lattice dynamics are more closely examined here. This study remains focused on the intermediate 
layers of the crust where the neutron superfluid fraction is predicted to be the most strongly suppressed.
To this end, fully three-dimensional band-structure calculations of the superfluid fraction 
in the limit of low temperatures and small velocities have been performed taking into account 
the quantum zero-point motions of ions. 
The model of  neutron-star crust and the calculations of the superfluid fraction are described in Sec.~\ref{sec:crust-model}. 
Numerical results are presented and discussed in Sec.~\ref{sec:results}. Concluding remarks are given in Sec.~\ref{sec:conclusion}.

\section{Microscopic model of dynamical neutron-star crust}
\label{sec:crust-model}

\subsection{Skyrme-Hartree-Fock equations in an dynamic lattice}

Let us consider the existence of an average stationary neutron superflow $\pmb{\bar V_{n,s}}$ in the crust frame. 
As in previous studies, $\bar V_{n,s}$ is assumed to be much smaller than Landau's velocity $V_{n,L}$ above which 
neutron Cooper pairs are broken (see, e.g., Ref.~\cite{allard2023} for a discussion in the nuclear context). 
In the weak-coupling approximation adopted here, the superfluid fraction can be evaluated from static 
calculations ignoring pairing~\cite{chamel2025}. The neutron single-particle wavefunction $\varphi(\rb,\sg)$ at 
position $\rb$ and spin coordinate $\sg$ in the volume $\Omega$ of the matter element under 
consideration, and the associated energies $\epsilon$ are determined from the Skyrme-Hartree-Fock (HF) equation~\cite{bender03}
($\hbar$ is the Planck-Dirac constant)
\beqy
\label{eq:HF}
-\pmb{\nabla}\cdot \frac{\hbar^2}{2m_n^{\oplus}(\rb)}\pmb{\nabla}\varphi(\rb,\sg) + U_n(\rb)\varphi(\rb,\sg) = 
\epsilon \varphi(\rb,\sg) \, ,
\eeqy
with Born - von K\'arm\'an periodic conditions on the boundaries of the volume $\Omega$. The spin-orbit coupling 
has been omitted since its contribution to the neutron superfluid density was shown to be negligible~\cite{chamel2005}.  
The fields are defined from functional derivatives of the Skyrme energy $E$ as follows
\begin{align}\label{eq:Fields1}
 &\frac{\hbar^2}{2m_n^{\oplus}(\rb)}=\frac{\delta E}{\delta \tau_n(\rb)}\equiv B_n(\rb)\,,  \qquad U_n(\rb)=\frac{\delta E}{\delta n_n(\rb)}\,,  
\end{align}
where 
\beqy\label{eq:local-density}
n_n(\rb)=\sum_{\sg=\pm 1} n_n(\rb,\sg;\rb,\sg)
\eeqy
is the local neutron number density,  and 
\beqy\label{eq:kinetic-density}
\tau_n(\rb)=\sum_{\sg=\pm 1}\int\text{d}^3\rp\; \delta (\rb-\rp) \pmb{\nabla}\cdot\pmb{\nabla^\prime} n_n(\rb,\sg;\rp,\sg)
\eeqy
is the neutron kinetic-energy density (in units of $\hbar^2/2m_n$, with $m_n$ the neutron mass) at position $\rb$. 
The neutron density matrix is defined as 
\begin{equation}
	n_n(\rb, \sg; \rp, \sg^\prime) = <\Psi\vert c_n(\rp,\sg^\prime)^\dagger c_n(\rb,\sg)\vert \Psi>\, ,
\end{equation}
where $c_n(\rb,\sg)^\dagger$ and $c_n(\rb,\sg)$ are respectively the creation and destruction operators for a neutron at position $\rb$ with spin projection $\sg=\pm 1$ (in units of $\hbar/2$) using the symbols $\dagger$ and $*$ for Hermitian and complex conjugations respectively,  and $\Psi$ denotes the many-body quantum state.  

Without any assumption about the distribution of ions, the neutron single-particle wave function 
can be quite generally expanded into plane waves
\beqy
\varphi(\rb,\sg)=\frac{1}{\sqrt{\Omega}}\sum_{\pmb{q}} \widetilde{\varphi}(\pmb{q}) \exp({\rm i} \pmb{q}\cdot\rb)\chi(\sg)
\eeqy
where the sum is over all possible wave vectors $\pmb{q}$, $\chi(\sg)$ is the Pauli spinor, and 
\beqy 
\sum_{\pmb{q}} |\widetilde{\varphi}(\pmb{q})|^2 = 1 \, .
\eeqy 
 
In Fourier space, the HF equation~\eqref{eq:HF} reads 
\beqy
\label{eq:eigenvalue-problem}
\sum_{\pmb{q'}} \biggl[ \pmb{q}\cdot \pmb{q'} \widetilde{B}_n(\pmb{q}-\pmb{q'})  + \widetilde{U}_n(\pmb{q}-\pmb{q'})\biggr] \widetilde{\varphi}(\pmb{q'}) = \epsilon\, \widetilde{\varphi}_n(\pmb{q})\, ,
\eeqy
where the Fourier coefficients are defined as 
\beqy
\widetilde{B}_n(\pmb{q})=\frac{1}{\Omega}\int{\rm d}^3r\, B_n(\rb) \exp(-{\rm i}\, \pmb{q}\cdot\rb)\, , 
\eeqy 
\beqy 
\widetilde{U}_n(\pmb{q})=\frac{1}{\Omega}\int{\rm d}^3r\, U_n(\rb) \exp(-{\rm i}\, \pmb{q}\cdot\rb)\, ,
\eeqy 
and the integrations are performed over the volume $\Omega$. 

At the densities of interest here, nuclear clusters are expected to be spherical. 
The fields arising from $N$ such ions located at position $\pmb{r_j}$ contained in the volume $\Omega$ are 
written as
\beqy
U_n(\rb)=\sum_{j=1}^{N} u_n(\rb-\pmb{r_j})\, , 
\eeqy
\beqy
B_n(\rb)=\sum_{j=1}^{N} b_n(\rb-\pmb{r_j})\, ,
\eeqy
where $u_n(\rb-\pmb{r_j})$ and $b_n(\rb-\pmb{r_j})$ are the local fields arising from the ion centered at $\pmb{r_j}$ defined by 
\beqy 
u(\rb-\pmb{r_j})=
\begin{cases}
U_{\rm WS}(\vert\rb-\pmb{r_j}\vert) - \dfrac{N-1}{N} U_{\rm WS}(R)  & \text{if} \ \vert\rb-\pmb{r_j}\vert < R\, ,\\
\dfrac{U_{\rm WS}(R)}{N}   & \text{if} \ \vert\rb-\pmb{r_j}\vert \geq R\, , 
\end{cases}
\eeqy 
\beqy 
b(\rb-\pmb{r_j})=
\begin{cases}
B_{\rm WS}(\vert\rb-\pmb{r_j}\vert) - \dfrac{N-1}{N} B_{\rm WS}(R) & \text{if} \ \vert\rb-\pmb{r_j}\vert < R\, ,\\
\dfrac{B_{\rm WS}(R)}{N} & \text{if} \ \vert\rb-\pmb{r_j}\vert \geq R\, .
\end{cases}
\eeqy 
Here $U_{\rm WS}(r)$ and $B_{\rm WS}(r)$ denote the corresponding fields inside a single spherical Wigner-Seitz cell of radius $R$ 
($r$ denoting in this case the radial coordinate as measured from the center of the cell). 
In this way, $U_n(\rb)$ and $B_n(\rb)$ are equal to $U_{\rm WS}(R)$ and $B_{\rm WS}(R)$ respectively in the interstitial region between clusters. 

The Fourier transforms of the mean fields are then given by 
\beqy
\widetilde{U}_n(\pmb{q})=S(\pmb{q}) \widetilde{U}_{\rm WS}(\pmb{q})-\dfrac{U_{\rm WS}(R)}{\Omega}\left\{ \left[1+(N-1)S(\pmb{q})\right]\int_{\rm cell} {\rm d}^3 r \exp{(-{\rm i} \pmb{q}\cdot\rb)} - \Omega \delta_{\pmb{q},\pmb{0}}\right\} \, ,
\eeqy
\beqy
\widetilde{B}_n(\pmb{q})=S(\pmb{q}) \widetilde{B}_{\rm WS}(\pmb{q})-\dfrac{B_{\rm WS}(R)}{\Omega}\left\{ \left[1+(N-1)S(\pmb{q})\right]\int_{\rm cell} {\rm d}^3 \rb \exp(-{\rm i} \pmb{q}\cdot\rb) - \Omega \delta_{\pmb{q},\pmb{0}}\right\} \, ,
\eeqy
where $\widetilde{U}_{\rm WS}(\pmb{q})$ and $\widetilde{B}_{\rm WS}(\pmb{q})$ are the Fourier transforms of the  fields in the spherical Wigner-Seitz cell 
of volume $\Omega_{\rm cell}=\Omega/N$, 
\beqy
\widetilde{U}_{\rm WS}(\pmb{q})=\frac{1}{\Omega_\textrm{cell}} \int_{\rm cell} {\rm d}^3 \rb\, \exp(-{\rm i} \pmb{q}\cdot\rb)  U_{\rm WS}(r)\, ,
\eeqy
and similarly for $\widetilde{B}_{\rm WS}(\pmb{q})$. All the information about the spatial distribution of ions is contained in the structure factor
\beqy\label{eq:structure-factor}
S(\pmb{q})=\frac{1}{N}\sum_{j=1}^N \exp(-{\rm i} \pmb{q}\cdot\pmb{r_j}) \, .
\eeqy

\subsection{Perfect rigid crystal}

If ions are arranged in a perfect static rigid lattice (as in cold atom experiments), their position vectors $\pmb{r_j}$  coincide with lattice translation vectors 
$\pmb{\ell_j}=n^j_1 \pmb{a_1}+n^j_2 \pmb{a_1}+n^j_3 \pmb{a_3}$, where $n^j_1$, $n^j_2$, $n^j_3$ are arbitrary integers, and $\pmb{a_1}$, $\pmb{a_2}$, $\pmb{a_2}$ are primitive basis vectors. According to the Floquet-Bloch theorem, (see, e.g., Ref.~\cite{kittel}), the single-particle states can be expressed as 
\beqy\label{eq:Bloch-state}
\varphi_{\alpha\pmb{k}}(\rb,\sg)=\frac{1}{\sqrt{\Omega}} \exp({\rm i} \pmb{k}\cdot\rb) \sum_{\beta} \widetilde{\varphi}_{\alpha\pmb{k}}(\pmb{G_\beta}) \exp({\rm i} \pmb{G_\beta}\cdot\rb)\chi(\sg)\, ,
\eeqy
where $\alpha$ is the band index, $\pmb{k}$ is the Bloch wave vector, and $\pmb{G_\alpha}=m^\alpha_1 \pmb{b_1}+m^\alpha_2 \pmb{b_1}+m^\alpha_3 \pmb{b_3}$ are reciprocal lattice vectors with $m^\alpha_1$, $m^\alpha_2$, $m^\alpha_3$ arbitrary integers (similarly for $\pmb{G_\beta}$),  and $\pmb{b_1}$, $\pmb{b_2}$, $\pmb{b_2}$ the corresponding primitive basis vectors  satisfying $\pmb{a_i}\cdot \pmb{b_j}=2\pi \delta_{ij}$ ($\delta_{ij}$ is Kronecker's symbol).  In such case, one only needs consider 
the matrix elements of $B_n(\rb)$ and $U_n(\rb)$ between plane wave states with wave vectors $\pmb{k}+\pmb{G_\alpha}$ and $\pmb{k}+\pmb{G_\beta}$, i.e. 
$\widetilde{B}_n(\pmb{G_\beta}-\pmb{G_\alpha})$ and $\widetilde{U}_n(\pmb{G_\beta}-\pmb{G_\alpha})$. For any reciprocal lattice vector $\pmb{G}$, one has $S(\pmb{G})=1$ and 
\beqy 
\int_{\rm cell} {\rm d}^3 \rb\, \exp(-{\rm i} \pmb{G}\cdot\rb) = \Omega_{\rm cell} \delta_{\pmb{G},\pmb{0}}\, .
\eeqy 
The Fourier transforms of the fields reduce to $\widetilde{U}_n(\pmb{G})= \widetilde{U}_{\rm WS}(\pmb{G})$ and 
$\widetilde{B}_n(\pmb{G})= \widetilde{B}_{\rm WS}(\pmb{G})$. The HF equation~\eqref{eq:eigenvalue-problem} thus reads
\beqy
\label{eq:HF-perfect-lattice}
\sum_{\beta} \biggl[(\pmb{k}+\pmb{G_\alpha})\cdot(\pmb{k}+\pmb{G_\beta}) \widetilde{B}_{\rm WS}(\pmb{G_\alpha}-\pmb{G_\beta})  + \widetilde{U}_{\rm WS}(\pmb{G_\alpha}-\pmb{G_\beta})\biggr] \widetilde{\varphi}_{\alpha\pmb{k}}(\pmb{G_\beta}) = \epsilon_{\alpha \pmb{k}}\, \widetilde{\varphi}_{\alpha\pmb{k}}(\pmb{G_\alpha})\, .
\eeqy

\subsection{Small vibrations of a perfect crystal}
\label{sec:small-vib}

Let us now consider small random displacements of each individual ion about their equilibrium position due to quantum 
zero-point motion, i.e. $\pmb{r_j}=\pmb{\ell_j}+\delta\rb(t)$ and $\vert\delta r(t)\vert \ll a$ ($a$ being the lattice spacing) 
is randomly fluctuating with time $t$. 
For a cubic lattice of pointlike ions in a rigid charge-compensating background of electrons, the thermal average of the structure factor~\eqref{eq:structure-factor} is given by 
(see, e.g., Appendix A of Ref.~\cite{kittel})
\beqy
\langle S(\pmb{q})\rangle=\exp\left(-\frac{1}{6} q^2\langle \delta \rb(t)^2\rangle\right)\, .
\eeqy
The disturbances are assumed to be sufficiently small that the single-particle states can still be described by 
Bloch wave functions~\eqref{eq:Bloch-state}. In such case, the HF equation~\eqref{eq:eigenvalue-problem} is formally the same as that for a rigid lattice~\eqref{eq:HF-perfect-lattice}, namely 
\beqy
\label{eq:HF-perfect-lattice+displacements}
\sum_{\beta} \biggl[(\pmb{k}+\pmb{G_\alpha})\cdot(\pmb{k}+\pmb{G_\beta}) \widetilde{B}_n(\pmb{G_\alpha}-\pmb{G_\beta})  + \widetilde{U}_n(\pmb{G_\alpha}-\pmb{G_\beta})\biggr] \widetilde{\varphi}_{\alpha\pmb{k}}(\pmb{G_\beta}) = \epsilon_{\alpha \pmb{k}}\, \widetilde{\varphi}_{\alpha\pmb{k}}(\pmb{G_\alpha})\, .
\eeqy
But now the Fourier coefficients are damped by the Debye-Waller factor: 
\beqy\label{eq:U+DW}
\widetilde{U}_n(\pmb{G})= \widetilde{U}_{\rm WS}(\pmb{G})\exp\left(-\frac{1}{6} G^2\langle \delta \rb(t)^2\rangle\right)\, ,
\eeqy
\beqy\label{eq:B+DW}
\widetilde{B}_n(\pmb{G})= \widetilde{B}_{\rm WS}(\pmb{G})\exp\left(-\frac{1}{6} G^2\langle \delta \rb(t)^2\rangle\right)\, .
\eeqy
Off-diagonal matrix elements of the mean fields in Eq.~\eqref{eq:HF-perfect-lattice} are thus exponentially suppressed by the motions of ions whereas the diagonal 
elements remain unchanged.

\subsection{Neutron superfluid density and effective ion mass}

The neutron superfluid density with quantum zero-point motion of ions can be calculated using the same expression as for a rigid lattice. 
Expanding the neutron mass current to linear order in $\bar V_{n,s}/V_{n,L}$ in the weak-coupling approximation, 
the neutron superfluid density is therefore given by~\cite{chamel2025}
\begin{eqnarray}\label{eq:super-density-weak2}
	\rho_{n,s} = \frac{m_n^2}{12\pi^3\hbar^2}\sum_\alpha \int_{\rm F} |\pmb{\nabla}_{\pmb{k}} \, 
	\epsilon_{\alpha\pmb{k}}|{\rm d}{\cal S}^{(\alpha)} \, ,
\end{eqnarray}
where the integration is performed over the Fermi surface. The single-particle energies $\epsilon$ therefore also $\rho_{n,s}$, however, 
now depends on the lattice vibrations through Eqs.~\eqref{eq:U+DW} and \eqref{eq:B+DW}. In turn, the ion fluctuations themselves are 
affected by the presence of superfluid neutrons~\cite{pethick2010}. This means that $\langle \delta \rb(t)^2\rangle$ depends on 
the superfluid density  $\rho_{n,s}$. The HF equation~\eqref{eq:HF-perfect-lattice+displacements} and the 
superfluid density~\eqref{eq:super-density-weak2} have therefore to be solved self-consistently. 

Ignoring the small mass difference between neutrons and protons, the bare ion mass gets dressed by the interactions with the surrounding 
neutrons. This effective ion mass can be written as $m_I^\star = m_n A^\star$ (the neutron mass is adopted here for the nucleon mass recalling that clusters are 
very neutron rich), where $A^\star$ represents the effective number of nucleons entrained 
by each ion and given by~\cite{chamelpagereddy2013}
\beqy \label{eq:Astar}
A^\star=A_{\rm cell}\left( 1- \frac{\rho_{n,s}}{\bar \rho}\right) \, .
\eeqy 
Here $A_{\rm cell}$ denotes the total number of neutrons and protons in the Wigner-Seitz cell and $\bar \rho$ is the average 
mass density. Since $\rho_{n,s} < \rho_{n,f}$, the number of entrained nucleons $A^\star > A$ is larger than the number 
of nucleons bound inside clusters defined by 
\beqy \label{eq:Abound}
A\equiv A_{\rm cell}\left( 1- \frac{\rho_{n,f}}{\bar \rho}\right) \, .
\eeqy  
From Eqs.~\eqref{eq:Astar} and \eqref{eq:Abound}, the effective ion mass can thus be directly expressed in terms of the neutron 
superfluid fraction as  
\beqy \label{eq:effective-ion-mass}
m^\star_I = m_I \frac{1- (\rho_{n,s}/ \rho_{n,f}) (\rho_{n,f}/\bar \rho)}{1- \rho_{n,f}/\bar \rho} \, .
\eeqy 
Since $\rho_{n,s}/\rho_{n,f}<1$, it follows that $m_I^\star>m_I$. 

As a consequence, the ion plasma frequency 
\beqy \label{eq:ion-plasma-frequency}
\omega^\star_I = \sqrt{\frac{4\pi Z^2 e^2 n_I}{m^\star_I}}\, ,
\eeqy 
is lowered~\cite{pethick2010,chamelpagereddy2013}, where $Z$ is the charge number of each ion, $e$ is the elementary electric charge, and $n_I=1/\Omega_{\rm cell}$ is the ion number density. At temperatures $T$ much lower than the ion plasma temperature $T^\star_{I}=\hbar \omega^\star_I/k_{\rm B}$ ($k_{\rm B}$ is Boltzmann's constant), 
the mean square displacement of ions is approximately given by (see, e.g., chapter 2 of Ref.~\cite{haensel2007})
\begin{equation}
\label{eq:zero-point-motion}
\langle \delta \rb(t)^2\rangle \approx  \frac{3 \hbar \langle \omega^\star_I/\omega_{ph}\rangle }{2 m^\star_I \omega^\star_I} \, , 
\end{equation}
where 
the dimensionless thermal average of the inverse phonon frequencies $\langle \omega^\star_I/\omega_{ph}\rangle$ depends on the 
lattice structure but not the ion mass. The ion motion is therefore damped by superfluid neutrons, $\langle \delta \rb(t)^2\rangle$ being 
reduced by a factor $\sqrt{m_I^\star/m_I}$.

\section{Results and discussion}
\label{sec:results}

\subsection{Numerical methods}

For numerical calculations, a finite number $\mathcal{N}^3$ of plane waves (corresponding to $\mathcal{N}$ reciprocal lattice vectors along each basis vector) is kept in the expansion~\eqref{eq:Bloch-state}. 
The HF equation~\eqref{eq:HF-perfect-lattice+displacements} is thus solved for each given Bloch wave vector $\pmb{k}$ in a spatial grid inside the primitive cell defined by the primitive basis vectors of the lattice, with $\mathcal{N}\times\mathcal{N}\times\mathcal{N}$ points defined by $\pmb{r}=(i_1/\mathcal{N})\pmb{a_1}+(i_2/\mathcal{N})\pmb{a_2}+(i_3/\mathcal{N})\pmb{a_3}$, where $i_1$, $i_2$, and $i_3$ are positive integers less or equal than $\mathcal{N}-1$. For a body-centered cubic (bcc) lattice, this corresponds to the following Cartesian coordinates (see Appendix~\ref{app:bcc}): 
\beqy 
x&=&\frac{a_{\rm bcc}}{2\mathcal N}\left( -i_1 +i_2 + i_3\right) \, ,\notag \\
y&=&\frac{a_{\rm bcc}}{2\mathcal N}\left( i_1 -i_2 + i_3\right) \, ,\notag \\
z&=&\frac{a_{\rm bcc}}{2\mathcal N}\left( i_1 +i_2 - i_3\right) \, .
\eeqy  
Calculations will also be carried out considering a face-centered cubic (fcc) lattice as this crystal structure might be more 
stable due to electron polarization~\cite{baiko2002}. In this case, the grid points are given by (see Appendix~\ref{app:fcc})
\beqy 
x&=&\frac{a_{\rm fcc}}{2\mathcal N}\left( i_2 + i_3\right) \, ,\notag \\
y&=&\frac{a_{\rm fcc}}{2\mathcal N}\left( i_1 + i_3\right) \, ,\notag \\
z&=&\frac{a_{\rm fcc}}{2\mathcal N}\left( i_1 +i_2\right) \, .
\eeqy 
Here $a_{\rm bcc}$ and $a_{\rm fcc}$ denote the size of the respective conventional cubic cell. 

Having computed the single-particle wavefunctions $\varphi_{\alpha\pmb{k}}(\rb,\sigma) $ and energies $\epsilon_{\alpha \pmb{k}}$, the gradients $\pmb{\nabla_k} \epsilon_{\alpha \pmb{k}}$ are calculated as~\cite{ChamelAllard2019}
\beqy
\label{eq:group-velocity}
\pmb{\nabla_k} \epsilon_{\alpha \pmb{k}}=\frac{-{\rm i} \hbar^2}{2}\sum_\sigma\int d^3\rb\,  \varphi_{\alpha\pmb{k}}(\rb,\sigma)^* \left[\frac{1}{m_n^\oplus(\rb)}\pmb{\nabla}+\pmb{\nabla}\frac{1}{m_n^\oplus(\rb)}\right]\varphi_{\alpha\pmb{k}}(\rb,\sigma) \, .
\eeqy
The neutron superfluid density $\rho_{n,s}$ is obtained by repeating the band-structure calculations for different $\pmb{k}$ and integrating Eq.~\eqref{eq:super-density-weak2} 
using the analytical method of Gilat and Raubenheimer~\cite{gilat1966} (see Ref.~\cite{chamel2025} for further details). 

\subsection{Numerical application}

For the sake of comparison with previous studies~\cite{chamel2025}, the fields $B_{\rm WS}(r)$ and $U_{\rm WS}(r)$ are taken from 
those of Ref.~\cite{onsi2008} calculated for the average baryon density $\bar n=0.03$~fm$^{-3}$ within the self-consistent 4th-order extended Thomas-Fermi (ETF) method~\cite{Brack_ea85} with proton shell 
correction added consistently via the Strutinsky integral (SI) theorem (the correction 
due to neutron-band structure was shown to be negligible~\cite{chamel2007}). This ETFSI method is 
a computationally very fast approximation to the self-consistent HF equations. The neutron chemical potential is determined using Eq.~(B21) of Ref.~\cite{pearson2012}. 

The spherical Wigner-Seitz cell of radius $R_{\rm cell}=23.30$ fm contains $A_{\rm cell}=1590$ nucleons, out of which $Z=40$ are protons. About 91\% of neutrons are free, and their average density is $n_{n,f}\approx 2.66 \times 10^{-2}$~fm$^{-3}$. From Eq.~\eqref{eq:Abound}, the number of nucleons bound in each ion is about $A\approx 181$. 
The lattice spacing is determined so that the volume $\Omega_{\rm cell}$ of the exact Wigner-Seitz cell is the same as that 
of the approximate spherical cell, namely $a_{\rm bcc}=R_{\rm cell}(8\pi/3)^{1/3}\approx 47.3$~fm for a bcc lattice, and  $a_{\rm fcc}=R_{\rm cell}(16\pi/3)^{1/3}\approx 59.6$~fm for an fcc lattice (see Appendix). Let us 
remark that computations for fcc crystals are significantly more costly than for bcc crystals as the lattice spacing is $2^{1/3}$ times larger. This means
that calculations for fcc lattice requires about twice as many grid points for the same grid spacing ($\delta x=\delta y=\delta z=a/(2\mathcal{N})$, where $a$ is the corresponding lattice spacing). Calculations have been performed in the primitive cell using a grid of $25\times 25\times 25=15625$ points for the bcc lattice, and $30\times 30\times 30=27000$ points for the fcc lattice, corresponding to a grid spacing $\delta x$ of about $0.95$~fm and $0.99$ fm respectively (the spacing along each primitive axis is respectively $\sqrt{3}$ and $\sqrt{2}$ times larger). In both cases, about 800 bands have been computed for each Bloch wave vector.

From Table 2.4 of Ref.~\cite{haensel2007}, the thermal average of the inverse phonon frequencies is given by  $\langle \omega_I/\omega_{ph}\rangle=2.79855$ for a bcc lattice, and $\langle \omega_I/\omega_{ph}\rangle=2.711982$ for an fcc lattice (see also Ref.~\cite{Baiko2001}). Initially, the effective ion mass is set to the bare ion mass $m_I$. Using Eq.~\eqref{eq:ion-plasma-frequency} with $m_I^\star=m_I$, the ion-plasma temperature is about $T_I\approx 4.13 \times 10^9$~K. This is much lower than the critical temperature of the neutron superfluid $T_{cn}\approx 1.05 \times 10^{10}$~K, as estimated from the BCS relation $k_{\rm B} T_{cn}=\Delta \exp(\gamma)/\pi$ ($\gamma\simeq 0.57722$ being the Euler-Mascheroni constant) with the realistic value $\Delta=1.59$~MeV for the $^1S_0$ pairing gap obtained from diagrammatic calculations in neutron matter at the same average neutron density taking into account both polarization and self-energy effects~\cite{cao2006}. The ion-plasma temperature still remains much higher than the typical temperature prevailing in the interior of observed isolated neutron stars (see, e.g., Ref.~\cite{potekhin2020}). The Debye-Waller factor entering Eqs.~\eqref{eq:U+DW} and \eqref{eq:B+DW} is calculated from Eq.~\eqref{eq:zero-point-motion} using Eq.~\eqref{eq:ion-plasma-frequency}. The corresponding root mean square displacement $\sqrt{\langle \delta \rb(t)^2\rangle }$ is about 1.65 fm for a bcc lattice and 1.63 fm for an fcc lattice. This is much smaller than the lattice spacing.  

Solving the HF equations~\eqref{eq:HF-perfect-lattice+displacements} for about 1400 Bloch wave vectors in the irreducible Brillouin zone (about 67000 in the full Brillouin zone) and computing the superfluid density~\eqref{eq:super-density-weak2} lead 
to a new estimate of the effective ion mass~\eqref{eq:effective-ion-mass}. The mean-square displacement is recalculated, and the HF equations are solved iteratively until the value of $\rho_{n,s}/\rho_{n,f}$ differs by less than three significant digits. Convergence is achieved after three iterations only. For comparison, calculations have been also performed ignoring the quantum zero-point motion of ions. Results are summarized in Tables~\ref{tab1} and \ref{tab2} for the bcc and fcc lattices respectively. 

As expected, the lattice dynamics increases the superfluid fraction, but the change is very small and lies within the errors of the numerical methods (see Ref.~\cite{chamel2025}). The influence of the crystal structure is more significant. In both cases, the effective ion mass is found to be substantially increased by about a factor 8 thus reducing the ion plasma temperature $T_I$ by a factor $2\sqrt{2}\approx 3$ (the temperature in known neutron stars still satisfies $T\ll T^\star_I$). The ions being effectively more massive, their motion is damped. Their root mean square displacement amounts to about 2\% of the lattice spacing. This justifies a posteriori the perturbative treatment of the ion motions and the low-temperature limit introduced in Sec.~\ref{sec:small-vib}.

\begin{table}
\begin{center}
\begin{tabular}{|c|c|c|c|}
\hline
&$\rho_{n,s}/\rho_{n,f}$ & $m^\star_I/m_I$ & $\sqrt{\langle \delta \rb(t)^2\rangle }/a_{\rm bcc}$ \\
\hline
           & 1                    & 1     & 0.034857 \\
\hline 
without DW & 0.079136             & 8.1897  &  0.020605 \\
\hline 
with DW & 0.081331                & 8.1725  &  0.020616 \\
        & 0.079522                & 8.1867  &  0.020607 \\
        & 0.079521                & 8.1867  &  0.020607\\
\hline
\end{tabular}
\caption{Properties of the inner crust of a neutron star with a bcc lattice structure at the average baryon number density $\bar n=0.03$ fm$^{-3}$: neutron superfluid fraction $\rho_{n,s}/\rho_{n,f}$, effective ion mass $m^\star_I/m_I$, and root mean square displacement $\sqrt{\langle \delta \rb(t)^2\rangle }$ of ions about their equilibrium position in units of the lattice spacing $a_{\rm bcc}$. The first line 
provides an estimate of the ion displacement ignoring the neutron superfluid dynamics. The second line shows results obtained for a rigid lattice, i.e. without the Debye-Waller (DW) factor. The last three lines show the iterative solutions taking into account the quantum zero-point motion of ions until self-consistency is achieved in the superfluid fraction up to three significant digits. See text for detail.} 
\label{tab1}
\end{center}
\end{table}

\begin{table}
\begin{center}
\begin{tabular}{|c|c|c|c|}
\hline
&$\rho_{n,s}/\rho_{n,f}$ & $m^\star_I/m_I$ & $\sqrt{\langle \delta \rb(t)^2\rangle }/a_{\rm fcc}$ \\
\hline
           & 1                       & 1       &   0.027234 \\
without DW & 0.068406                & 8.2734  &   0.016313 \\
\hline 
with DW    & 0.071351                & 8.2504  &   0.016324 \\
           & 0.069492                & 8.2650  &   0.016317 \\
           & 0.069490                & 8.2650  &   0.016317  \\
\hline
\end{tabular}
\caption{same as Table~\ref{tab1} for an fcc lattice.} 
\label{tab2}
\end{center}
\end{table}

\section{Conclusion}
\label{sec:conclusion}

The assumption of a rigid lattice in the calculations of the neutron superfluid dynamics in the inner crust 
of a neutron star has been critically examined. To this end, the neutron superfluid fraction has been calculated 
in the limit of small neutron superfluid velocities $\bar V_{n,s}\ll V_{n,L}$ in the weak-coupling approximation. 
The quantum zero-point motion of ions about their equilibrium position has been included through the introduction 
of the Debye-Waller factor in the Fourier transforms of the mean fields. 

Full three-dimensional band-structure calculations have been performed in the intermediate region of the inner crust 
at the average baryon density 0.03~fm$^{-3}$. The suppression of the superfluid fraction effectively increases the mass of 
the ions, thus damping their displacement. These effects have been computed self-consistently, considering both
bcc and fcc crystal structures. In both cases, the superfluid fraction $\rho_{n,s}/\rho_{n,f}\approx 8\%$ remains 
essentially unchanged. Treating the crust as a perfect crystal is therefore a very good approximation in these layers. 
This is because the effective mass of the ions is dramatically increased by the presence of the neutron superfluid. 
These results challenge the interpretation of pulsar glitches~\cite{andersson2012,chamel2013,delsate2016} and may 
have important implications for the cooling of neutron stars~\cite{chamelpagereddy2013}. 

The influence of the lattice dynamics on the superfluid fraction may be more significant in deeper regions of the 
neutron-star crust, especially in the nuclear pasta mantle which has been the focus of most studies in recent years~\cite{kashiwaba2019,sekizawa2022,minami2022,kenta2024,almirante2024,almirante2024b}. This calls for further 
investigations.

\appendix

\section{Crystal structures}
\subsection{Body-centered cubic lattice}
\label{app:bcc}

The primitive basis vectors of a body-centered cubic lattice are given by 
\beqy \label{eq:bcc-primitive-vectors}
 \pmb{a_1}&=&\frac{a}{2}\left( -\pmb{\hat x} + \pmb{\hat y} + \pmb{\hat z}\right) \, ,\notag \\ 
 \pmb{a_2}&=&\frac{a}{2}\left( \pmb{\hat x} - \pmb{\hat y} + \pmb{\hat z}\right) \, ,\notag \\ 
 \pmb{a_3}&=&\frac{a}{2}\left( \pmb{\hat x} + \pmb{\hat y} - \pmb{\hat z}\right) \, ,
\eeqy  
where $\pmb{\hat x}$, $\pmb{\hat y}$, $\pmb{\hat z}$ are the Cartesian unit vectors and $a$ is the 
size of the conventional cubic cell. The volume of a primitive cell is therefore 
$\Omega_{\rm cell}=|\pmb{a_1}\cdot \pmb{a_2}\times \pmb{a_3}|=a^3/2$. 

The reciprocal lattice is face-centered cubic with basis vectors
\beqy \label{eq:bcc-reciprocal-primitive-vectors}
 \pmb{b_1}&=&\frac{2\pi}{a}\left(\pmb{\hat y} + \pmb{\hat z}\right) \, ,\notag \\ 
 \pmb{b_2}&=&\frac{2\pi}{a}\left( \pmb{\hat x}  + \pmb{\hat z}\right) \, ,\notag \\ 
 \pmb{b_3}&=&\frac{2\pi}{a}\left( \pmb{\hat x} + \pmb{\hat y} \right) \, .
\eeqy  

\subsection{Face-centered cubic lattice}
\label{app:fcc}

The corresponding primitive basis vectors of a face-centered cubic lattice are given by 
\beqy \label{eq:fcc-primitive-vectors}
 \pmb{a_1}&=&\frac{a}{2}\left(  \pmb{\hat y} + \pmb{\hat z}\right) \, ,\notag \\ 
 \pmb{a_2}&=&\frac{a}{2}\left( \pmb{\hat x} + \pmb{\hat z}\right) \, ,\notag \\ 
 \pmb{a_3}&=&\frac{a}{2}\left( \pmb{\hat x} + \pmb{\hat y} \right) \, .
\eeqy  
The volume of a primitive cell is therefore 
$\Omega_{\rm cell}=|\pmb{a_1}\cdot \pmb{a_2}\times \pmb{a_3}|=a^3/4$. 

The reciprocal lattice is a body-centered cubic with basis vectors
\beqy \label{eq:fcc-reciprocal-primitive-vectors}
 \pmb{b_1}&=&\frac{2\pi}{a}\left(- \pmb{\hat x}+\pmb{\hat y} + \pmb{\hat z}\right) \, ,\notag \\ 
 \pmb{b_2}&=&\frac{2\pi}{a}\left( \pmb{\hat x} -\pmb{\hat y} + \pmb{\hat z}\right) \, ,\notag \\ 
 \pmb{b_3}&=&\frac{2\pi}{a}\left( \pmb{\hat x} + \pmb{\hat y} - \pmb{\hat z} \right) \, .
\eeqy  

\begin{acknowledgments}
This work was financially supported by Fonds de la Recherche Scientifique (Belgium) under Grant No. PDR T.004320.  The author is a member of BLU-ULB (Brussels Laboratory of the Universe, blu.ulb.be).
\end{acknowledgments}

\bibliography{references.bib}
\end{document}